\documentclass[a4paper,twocolumn,11pt,accepted=2020-10-28]{quantumarticle}
\pdfoutput=1
\usepackage[utf8]{inputenc}
\usepackage[english]{babel}
\usepackage[T1]{fontenc}
\usepackage{amsmath}
\usepackage{hyperref}

\usepackage{tikz}
\usepackage{lipsum}

\usepackage{quoting}
\usepackage[utf8]{inputenc} 
\usepackage[english]{babel}
\usepackage{graphicx}
\usepackage{epstopdf}
\usepackage{hyphenat}
\usepackage{amsmath}
\usepackage{wrapfig}
\usepackage{setspace}
\usepackage{hyperref}
\usepackage{textcomp}
\usepackage{subfigure}
\usepackage{rotating}
\usepackage{wrapfig}
\usepackage{braket}
\usepackage{amssymb}
\usepackage{amsmath}
\usepackage{xcolor}
\usepackage{bbold}
\allowdisplaybreaks

\begin{document}

\title{Time Observables in a Timeless Universe}

\author{Tommaso Favalli}
\affiliation{QSTAR, INO-CNR and LENS, Largo Enrico Fermi 2, I-50125 Firenze, Italy}
\affiliation{Universit\'a degli Studi di Napoli Federico II, Via Cinthia 21, I-80126 Napoli, Italy}
\email{favalli@lens.unifi.it}
\orcid{0000-0001-7952-5835}
\author{Augusto Smerzi}
\affiliation{QSTAR, INO-CNR and LENS, Largo Enrico Fermi 2, I-50125 Firenze, Italy}
\email{augusto.smerzi@ino.it}
\orcid{0000-0002-4967-6939}

\maketitle

\begin{abstract}
 Time in quantum mechanics is peculiar: it is an observable that cannot be associated to an Hermitian operator. As a consequence it is impossible to explain dynamics in an isolated system without invoking an external classical clock, a fact that becomes particularly problematic in the context of quantum gravity. 
 An unconventional solution was pioneered by Page and Wootters (PaW) in 1983. PaW showed that dynamics can be an emergent property of the entanglement between two subsystems of a static Universe.  
 In this work we first investigate the possibility to introduce in this framework a Hermitian time operator complement of a clock Hamiltonian having an equally-spaced energy spectrum. An Hermitian operator complement of such Hamiltonian was introduced by Pegg in 1998, who named it \lq\lq Age\rq\rq.  
 We show here that Age, when introduced in the PaW context, can be interpreted as a proper Hermitian time operator conjugate to a \lq\lq good\rq\rq clock Hamiltonian. We therefore show that, still following Pegg's formalism, it is possible to introduce in the PaW framework bounded clock Hamiltonians with an unequally-spaced energy spectrum with rational energy ratios. In this case time is described by a POVM and we demonstrate that Pegg's POVM states provide a consistent dynamical evolution of the system even if they are not orthogonal, and therefore partially un-distinguishables.
  
\end{abstract}


\section{Introduction}
\label{Introduction}

\subsection{Time and Clocks}
\label{TimeObservables}
Observables in quantum theory are represented by Hermitian operators with the exception of time \cite{pauli,bush} 
(and of a few more observables, including phases \cite{phase}). In Quantum Mechanics, as 
in Newtonian physics, time is an absolute ``external'' real valued parameter
that flows continuously, independently from the material world. 
A change of perspective 
from this abstract Newtonian concept was introduced in the theory of Relativity. Here time is an ``internal'' degree of freedom 
of the theory itself, operationally defined by  
``what it is shown on a clock'', with the clock being a wisely chosen physical system \cite{rovelli}.

An interesting question is if an operational approach would also be possible in
Quantum Mechanics by considering time as ``what it is shown in a quantum clock'' \cite{pagewootters}.
A proposal was formulated in 1983 by Don N. Page and William K. Wootters (PaW) \cite{pagewootters,wootters}. Motivated by 
``the problem of time''
in canonical quantization of gravity (see for example \cite{dewitt,isham}) and considering a ``Universe'' in 
a stationary global state satisfying the Wheeler-DeWitt equation
$\hat{H}\ket{\Psi} = 0$, 
PaW suggested that dynamics can be considered as 
an emergent property of entangled subsystems, with 
the clock provided by a quantum spin rotating under the action of an applied magnetic field.
This approach has recently attracted a large interest and has stimulated several extensions and generalisations (see for example \cite{lloydmaccone,esp2,vedral,vedraltemperature,macconeoptempoarrivo,wigner,peggtimeqm,interacting,cuccoli}), including an experimental illustration \cite{esp1}. A brief summary of the Page and Wootters theory is given in Appendix A.  
\subsection{Time Observables in a Timeless Quantum System}
\label{TimeObservables2}
We consider here an isolated non-relativistic quantum system that in the following we call ``Universe''.
The Hilbert space of the ``Universe'' is composed by  
a ``clock-subspace'' $C$ that keeps track of time
and the ``system-subspace'' $S$ of the rest of the Universe.
In absence of an external temporal reference frame
we write the Schrödinger equation of the Universe as: 
\begin{equation}\label{sc}
(\hat{H}_s - i \hbar \frac{\partial}{\partial t_c} )\ket{\Psi} = 0
\end{equation}
The first term $\hat{H}_s$ is the Hamiltonian of the sub-system $S$. We interpret the second term 
as a (possibly approximate) time representation of the clock-subspace Hamiltonian:
\begin{equation}\label{ch}
- i \hbar \frac{\partial}{\partial t_c} \to \hat{H}_c.
\end{equation}
Under the implicit assumption that the two subsystems $C$ and $S$ are not interacting, 
Eq. (\ref{sc}) becomes: 
\begin{equation} \label{scm}
(\hat{H}_s + \hat{H}_c)\ket{\Psi} = 0.
\end{equation}
The time representation of the clock Hamiltonian Eq. (\ref{ch}) would be correct, $- i \hbar \frac{\partial}{\partial t_c} = \hat{H}_c$,
only if $\hat{H}_c$ has a continuous, unbounded spectrum. 
In this case we could write the Hermitian time operator in
the energy representation as $\hat T = - i \hbar \frac{\partial}{\partial E_c}$.
Since we consider here an isolated physical system 
of finite size, the introduction of unbounded Hamiltonians with a continuous spectrum would not be possible.
As a consequence, an Hermitian time operator written in differential form as in Eq. (\ref{sc})
cannot be introduced within standard approaches \cite{pauli}.


In this work we explore the possibility to construct a time Hermitian observable that is conjugate to a clock Hamiltonian having an
eigenvalue spectrum with a finite lower bound. It is clear, as already mentioned, that such exploration would be fruitless for the
simple reason that its existence would contradict the Stone-von Neumann theorem. 
A way out was considered by D. T. Pegg \cite{pegg} (see also \cite{peggbar})
who suggested a protocol to construct an Hermitian operator, named \lq\lq Age\rq\rq, complement of a lower-bounded Hamiltonian with equally-spaced energy eigenvalues.
The idea was to consider an Hamiltonian with an energy cut-off, calculate all quantities of interest, and eventually 
remove the cut-off by letting go to infinity the upper bound. 

Age is conjugate to the Hamiltonian in the sense that it is the generator of energy shifts while the Hamiltonian is the generator of translations of 
the eigenvalues of Age. However, as Pegg emphasized, and consistently with the Pauli objection \cite{pauli},
the Age operator cannot be considered,
as a \textit{bona-fide} time operator. This because
Age is a property of the system itself and crucially depends on its state. In particular, while
for particular states the rate of change of Age's mean value can be constant,
for an energy eigenstate its mean value does not evolve: {\it a system in a stationary state would not age as time goes on} \cite{pegg}.
Pegg also considered a larger set of Hamiltonians having unequally-spaced energy eigenvalues. Also in this case it is possible to write down
a complement of the Hamiltonian whose eigenvalues can be used to construct a probability-operator measure (POVM).

The central result of our work is to show that the Pegg formalism finds a sound physical interpretation
when incorporated in the Page-Wooters framework. We introduce as ``good'' clock a device described by an Hamiltonian having
equally spaced eigenvalues. In this case time is described by an Hermitian operator. A device governed by an Hamiltonian having 
rational energy differences can still provide a good clock mathematically described by a POVM. We show that in both cases we 
recover the Schr\"odinger dynamical evolution of the system $S$.

\section{Time from Entanglement}
\label{Time from Entanglement}


\subsection{The Clock Subspace}
\label{TheClockSubspace}
The first problem to deal with is the introduction of a good clock.
We define as good clock a physical system governed by a lower-bounded Hamiltonian having discrete, equally-spaced, energy levels, see also 
\cite{simile,simile2}
(a generalisation to non-equally spaced levels will be discussed in Section IV): 
\begin{equation}
\hat{H}_c = \sum_{n=0}^{s} E_n \ket{E_n}\bra{E_n},
\end{equation}
where $s+1$ is the dimension of the clock space that, following D. T. Pegg \cite{pegg}, we first consider as finite.
We now search for an Hermitian observable $\hat{\tau}$ in the clock space that is conjugated to the clock Hamiltonian $\hat{H}_c$.
We define the time states (we set $\hbar =1$)
\begin{equation}\label{taumautovet}
\ket{\tau_m}_c = \frac{1}{\sqrt{s+1}} \sum_{n=0}^{s} e^{-i E_n \tau_m} \ket{E_n}_c 
\end{equation}
with $\tau_m = \tau_0 + m\frac{T}{s+1}$, $E_n = E_0 + n\frac{2\pi}{T}$ and $m,n=0,1,...,s$.
Eq. (\ref{taumautovet}) provides an orthonormal and complete basis since 
\begin{equation}
\braket{\tau_m|\tau_{m'}}=\delta_{m,m'}
\end{equation}
and
\begin{equation}\label{identity}
\sum_{m=0}^{s} \ket{\tau_m}\bra{\tau_m} = \mathbb{1}_c.
\end{equation}

With the states (\ref{taumautovet}) we can define the Hermitian time operator
\begin{equation}\label{tauop}
\hat{\tau} = \sum_{m=0}^{s} \tau_m \ket{\tau_m}\bra{\tau_m}
\end{equation}
that is conjugated to the Hamiltonian $\hat{H}_c$. It is indeed easy to show
that $\hat{H}_c$ is the generator of shifts in $\tau_m$ values and, viceversa, $\hat{\tau}$ is the generator of energy shifts:
\begin{equation}\label{property11}
\ket{\tau_m}_c = e^{-i \hat{H}_c(\tau_m - \tau_0)}\ket{\tau_0}_c
\end{equation}
and
\begin{equation}
\ket{E_n}_c = e^{i  \hat{\tau} (E_n - E_0)} \ket{E_0}_c	.
\end{equation}
A second important property of the clock states is their ciclic condition: $\ket{\tau_{m=s+1}}=\ket{\tau_{m=0}}$.
The time taken by the system to return to its initial state is 
\begin{equation}\label{1c}
T=\frac{2\pi}{\delta E}
\end{equation}
with $\delta E$ being the spacing between two neighbouring energy eigenvalues. Conversely, the smallest time interval is
\begin{equation}\label{2c}
\delta\tau= \tau_{m+1} -\tau_{m} = \frac{2\pi}{\delta E \left(s+1\right)} .
\end{equation}
To summarise: the greater is the spectrum of the clock Hamiltonian, the smaller is the spacing $\delta \tau$ between two eigenvalues of the clock. 
The smaller is the distance between two eigenvalues of the clock energy, the larger the range $T$ of the eigenvalues $\tau_m$.
The final crucial step is to choose the value of $s$. Following Pegg's prescription \cite{pegg},  
this has to be first taken finite in order to allow the calculation of
all quantities of interest, including the Schr\"odinger equation, that will therefore functionally depend on $s$.
The physical values of the observables are eventually obtained in the limit $s \longrightarrow \infty$.
Obviously, this limit implies a continuous flow of time, but nothing forbids, in principle, to choose $s$ large but finite so to preserve
a discrete time evolution. The two prescriptions would give different predictions for measured values of observables.
\subsection{Dynamics}
\label{EvolutionSingleState}
We consider the total Hilbert space of the Universe $\mathcal{H} = \mathcal{H}_c \otimes \mathcal{H}_s$, with $\mathcal{H}_c$ and $\mathcal{H}_s$ 
having dimension $d_c = s+1$ and $d_s$ respectively. We require that our  
``good clock'' has $d_c \gg d_s$. 
A general bipartite state of the Universe can be written as
\begin{equation}
\ket{\Psi} = \sum_{n=0}^{d_c-1} \sum_{k=0}^{d_s-1} c_{n,k} \ket{E_n}_c \otimes \ket{E_k}_s.
\end{equation}
We impose the constraint Eq. (\ref{scm}) $\hat{H}\ket{\Psi}=0$ and, under the assumption that 
the spectrum of the clock Hamiltonian is sufficiently dense (namely, that to each energy state of the system $S$ there is a state of the clock for which Eq. (\ref{scm}) is satisfied),
we obtain for the state of the Universe
\begin{equation}
\ket{\Psi} = \sum_{k=0}^{d_s-1} \tilde{c}_{k} \ket{E=-E_k}_c \otimes \ket{E_k}_s
\end{equation}
with $\sum_{k} \left|\tilde{c}_{k} \right|^2=1$. With the resolution of the identity (\ref{identity}), we write
\begin{equation}\label{stato1}
\ket{\Psi} = 
\frac{1}{\sqrt{d_c}} \sum_{m=0}^{d_c-1} \ket{\tau_m}_c \otimes  \sum_{k=0}^{d_s-1}\tilde{c}_{k} e^{-iE_k\tau_m}\ket{E_k}_s.
\end{equation}
By writing a generic state of the system as
$\ket{\phi_{m}}_s= \sum_{k=0}^{d_s-1}\tilde{c}_{k} e^{-iE_k\tau_m}\ket{E_k}_s$, the state (\ref{stato1}) becomes
\begin{equation}\label{stato2}
\ket{\Psi}  =\frac{1}{\sqrt{d_c}} \sum_{m=0}^{d_c-1} \ket{\tau_m}_c \otimes \ket{\phi_m}_s .  
\end{equation}
It is interesting to note, and we emphasise, that the state $\ket{\phi_{m}}_s$ is related to the the global $\ket{\Psi}$ of the Universe by 
\begin{equation}\label{definition}
\ket{\phi_{m}}_s = \frac{\braket{\tau_m|\Psi}}{1/\sqrt{d_c}} 
\end{equation}
that is the Everett \textit{relative state} definition of the subsystem $S$ with respect to the clock system $C$ \cite{everett}. As
pointed out in \cite{vedral}, this kind of projection has nothing to do with a measurement. Rather, $\ket{\phi_m}_s$ 
is a state of $S$ conditioned to the clock $C$ in the state $\ket{\tau_{m}}_c$.

Now, following the PaW framework and using Eq. (\ref{definition}), the constraints Eq. (\ref{scm}) and 
Eq. (\ref{property11}), we have:
\begin{equation}\label{ev}
\begin{split}
\ket{\phi_{m}}_s 
&= \sqrt{d_c}\bra{\tau_0}e^{i\hat{H}_c(\tau_m-\tau_0)}\ket{\Psi}=\\ \\&= \sqrt{d_c}\bra{\tau_0}e^{i(\hat{H}-\hat{H}_s)(\tau_m-\tau_0)}\ket{\Psi}=
\\ \\&=
e^{-i\hat{H}_s (\tau_{m}-\tau_0)}\ket{\phi_{0}}_s  
\end{split}
\end{equation}
where $\ket{\phi_0}_s =\frac{\braket{\tau_0|\Psi}}{1/\sqrt{d_c}}  = \sum_{k=0}^{d_s-1} \tilde{c}_k  e^{- i E_k \tau_0} \ket{E_k}_s$.
The Eq. (\ref{ev}) provides the Schrödinger evolution of $S$ with respect to the clock time. 

Now we can also consider the global state  written in the form (\ref{stato2}) and, through (\ref{ev}), we can consider the unitary operator $\hat{U}_{s} (\tau_{m}-\tau_0) = e^{-i \hat{H}_s (\tau_{m}-\tau_0)}$ \cite{lloydmaccone}. With this choice the state of the global system can be written as 
\begin{equation}
\begin{split}
\ket{\Psi}  =\frac{1}{\sqrt{d_c}} \sum_{m=0}^{d_c-1} \ket{\tau_m}_c \otimes \hat{U}_{s} (\tau_{m}-\tau_0) \ket{\phi_0}_s
\end{split}
\end{equation}
where explicitly is included the entire time history of the Universe.
We conclude this Section by noticing that 
the conditional probability of obtaininig the outcome $a$ for the system $S$ when measuring the observable $A$ at a certain time $\tau_{m}$ is given, as expected, by the Born rule:

\begin{equation}
\begin{split}
P(a\: on\: S \: |\: \tau_m \: on \:C) &=\frac{P(a \: on \: S , \: \tau_m \: on \: C)}{P( \tau_m \: on \: C)}  =\\ \\&=  \left|	\bra{a}\hat{U}_{s} (\tau_{m}-\tau_0) \ket{\phi_0}\right|^2  .
\end{split}
\end{equation}

\section{The Hermitian Time Operator}
\label{TAUasTimeOperator}
Here we show that within the PaW framework the operator $\hat{\tau}$ has the expected properties of a Hermitian time observable. 
It is well known that Pauli objected about the existence of a time Hermitian operator because time is continuous and 
unbounded in the past and in the future while general Hamiltonians have a lower bounded (continuous or discrete) spectrum \cite{leonmaccone}.
Pegg's Age operator overcome the energy objection \cite{pauli} since $\hat{\tau}$ has a discrete spectrum and cyclical boundary conditions
while the appropriate limits are taken only after calculating whatever of interest. 
The question we address here is why $\hat{\tau}$ can not be considered as a proper time operator outside the PaW mechanism.
As clearly pointed out by Pegg, $\hat{\tau}$ has dimensions of time but it is a property of the quantum system, and it strongly depends on the state of the system. With a quantum system with Hamiltonian $\hat{H}$ we would be forced to consider $\hat{\tau}$ defined 
on the space of the system itself. So the evolution of the mean value of $\hat{\tau}$ operator with respect to an external time has to be constant or at least not zero, otherwise the dynamics would freeze:
\begin{equation}\label{mammamia}
\begin{split}
\frac{d \langle \hat{\tau} \rangle}{dt} &= - i \bra{\psi} \left[ \hat{\tau}, \hat{H} \right]\ket{\psi} \\ \\&\propto \sum_{n, n'} (E_{n'}-E_n)\braket{\psi|E_{n'}} \braket{E_n|\psi}
\end{split}
\end{equation}
where $\ket{\psi}$ is a generic state of the system .
If we consider the system in an energy eigenstate (that is $\ket{\psi}=\ket{E_i}$), we obtain
\begin{equation}\label{mammamiamiamia}
\frac{d \langle \hat{\tau} \rangle}{dt} = 0 
\end{equation}
which means that the $\tau_{m}$ values stops running over time. So, outside the PaW framework, the $\hat{\tau}$ operator can not be considered as a time observable, but as a property of the system that has dimension of time.
Conversely, within the PaW framework, we have a global stationary state that includes the whole time history of $S$ with respect to $C$.
%
An energy eigenstate of the system $S$ evolves with an unobservable global phase
%
\begin{equation}\label{dausareperordinaryqm}
\ket{\phi_m}_s=e^{-i  E_k \tau_{m}}\ket{E_k}_s .
\end{equation}
However, this does not mean that the Universe stops. Indeed, from the fact that in the clock space $\hat{\tau}$ and $\hat{H}_c$ are conjugated operators, it follows that, even if taking the system $S$ in an energy eigenstate $\ket{E_k}$ forces the clock in an eigenstate of $\hat{H}_c$, all time states exist (indeed, thanks to the fact that $\hat{\tau}$ and $\hat{H}_c$ are incompatible observables, for construction we have $\ket{E_k} \propto \sum_{m} e^{-iE_k \tau_{m}} \ket{\tau_m}$). 
The $\hat{\tau}$ operator that Pegg's defined as complement of the Hamiltonian becomes a proper time operator when included
in the PaW framework. This happens in general with any choice of the clock Hamiltonian, as discussed by Leon and Maccone in [23],
because in the Page and Wootter theory the concept of external time is eliminated (or in any case becomes irrelevant), and time is an emerging property of entanglement between the system $S$ and the clock $C$ imposed by 
the Wheeler-DeWitt constraint.

\section{Unequally Spaced Energy Levels for the Clock Hamiltonian}
\label{unequally}
With the perspective to extend the set of Hamiltonians useful to describe a clock, we now consider the case in which the clock Hamiltonian does not have equally spaced energy levels, but non-degenerate eigenstates having rational energy differences. 
In this case we cannot define an Hermitian operator but we can still 
introduce a probability-operator measure, complement of such Hamiltonian \cite{pegg}.
\subsection{Discrete Flow of Time}
\label{discreteflow}
We consider a quantum system described by $p+1$ energy states $\ket{E_i}$ and $E_i$ energy levels with $i=0,1,2,...,p$ such that
\begin{equation}\label{numraz}
\frac{E_i - E_0}{E_1 - E_0} = \frac{C_i}{B_i},
\end{equation}
where $C_i$ and $B_i$ are integers with no common factors. We can write
\begin{equation}\label{ei}
E_i = E_0 + r_i \frac{2\pi}{T}
\end{equation}
where $T=\frac{2\pi r_1}{E_1 - E_0}$, $r_i = r_1\frac{C_i}{B_i}$ for $i>1$ (with $r_0=0$) and $r_1$ 
equal to the lowest common multiple of the values of $B_i$.

In this space we define the states 
\begin{equation}\label{defstatialpha}
\ket{\alpha_m}_c  = \frac{1}{\sqrt{d_c}}\sum_{i=0}^{d_c-1}e^{-i E_i \alpha_m}\ket{E_i}_c
\end{equation}
where $d_c = p+1$ and
\begin{equation}\label{alpha1}
\alpha_m = \alpha_0 + m \frac{T}{s+1} 
\end{equation} 
with $m=0,1,2,...,s$ and $s+1=D \ge r_p$. The number of $\ket{\alpha_m}$ states is therefore greater than the number of energy states in $\mathcal{H}_c$ and the $s+1$ values of $\alpha_m$ are uniformly distributed over $T$.
The resolution of the identity (\ref{identity}) is now replaced by (see Appendix B)
\begin{equation}\label{pomidentity}
\mathbb{1}_c = \frac{p+1}{s+1} \sum_{m=0}^{s} \ket{\alpha_{m}}\bra{\alpha_{m}}.
\end{equation}
As in the previous discussion, we can now consider a general state in the space $\mathcal{H}=\mathcal{H}_c \otimes \mathcal{H}_s$ and require that it satisfies the PaW constraint. 
By writing 
\begin{equation}
\ket{\Psi} = \sum_{n=0}^{d_c-1} \sum_{k=0}^{d_s-1} c_{n,k} \ket{E_n}_c \otimes \ket{E_k}_s
\end{equation}
and imposing $\hat{H}\ket{\Psi} = 0$ (considering $d_c \gg d_s$), we obtain
\begin{equation}
\ket{\Psi} = \sum_{k=0}^{d_s-1} \tilde{c}_k \ket{E=- E_k}_c \otimes \ket{E_k}_s
\end{equation}
with $\sum_{k=0}^{d_s-1} \left| \tilde{c}_k \right|^{2} = 1$. We can now apply the resolution of the identity (\ref{pomidentity}) to the state $\ket{\Psi}$ and obtain $(D=s+1)$:
\begin{equation}\label{applicazionerisoluzioneidentita}
\begin{split}
\ket{\Psi} & = \frac{d_c}{D} \sum_{m=0}^{D-1} \ket{\alpha_{m}}\braket{\alpha_{m}|\Psi} = \\ \\&= \frac{\sqrt{d_c}}{D} \sum_{m=0}^{D-1} \ket{\alpha_{m}}_c \otimes \sum_{k=0}^{d_s-1} \tilde{c}_k e^{-i E_k \alpha_{m}} \ket{E_k}_s . 
\end{split}
\end{equation}
We notice here that the states $\ket{\alpha_{m}}_c$ are not orthogonal. This introduces a possible conceptual warning that 
needs to be discussed. It is clear that by considering time states that are not orthogonal implies that these are
partially indistinguishable with a single measurement, the probability of indistinguishability being proportional to 
$| \braket{\alpha_{m'}|\alpha_m}|^2$.
%
The partial indistinguishability of the states $\ket{\alpha_{m}}_c$ implies an overlap between different times. 
We show, however, that also in this case the time evolution is described by the Schr\"odinger equation. Considering
\begin{equation}\label{hfhfhfhfhfhfh}
\ket{\phi_m}_s =  \frac{\braket{\alpha_m|\Psi}}{1/\sqrt{d_c}} ,
\end{equation}
we obtain $\ket{\phi_m}_s = \sum_{k=0}^{d_s-1} \tilde{c}_k e^{-i \alpha_{m} E_k} \ket{E_k}_s$ for the state of the system $S$ (see Appendix C). 
Therefore, even if time states are partially indistinguishable, the state of the system $S$, conditioned on a given $\ket{\alpha_{m}}_c$,
evolves with $\alpha_m$. Indeed, thanks to the fact that 
\begin{equation}
\ket{\alpha_{m}}_c = e^{-i\hat{H}_c (\alpha_{m}-\alpha_0)} \ket{\alpha_{0}}_c,
\end{equation}
using once again the constraint (\ref{scm}) and the (\ref{hfhfhfhfhfhfh}), we obtain
\begin{equation}\label{evoluzionealpha}
\ket{\phi_{m}}_s = e^{-i \hat{H}_s (\alpha_{m}-\alpha_{0})} \ket{\phi_0}_s
\end{equation}
that is the Schrödinger evolution for the state $\ket{\phi_m}_s$ with the Hamiltonian $\hat{H}_s$.
We notice here that POVMs generalizing the one introduced by Pegg 
have been discussed in \cite{review}. The consequence of using POVM is that time states are not fully indistinguishable. This suggests a possible extension of the definition of the Everett relative states where is still possible to have
a consistent dynamical evolution of the system $S$. 

To conclude this Section we notice that, through equation (\ref{evoluzionealpha}), we can again define  the unitary operator $\hat{U}_{s} (\alpha_{m}-\alpha_0) = e^{-i \hat{H}_s (\alpha_{m}-\alpha_0)}$. With this choice the state of the global system can be written as 
\begin{equation}
\begin{split}
\ket{\Psi}  =\frac{\sqrt{d_c}}{D} \sum_{m=0}^{D-1} \ket{\alpha_m}_c \otimes \hat{U}_{s} (\alpha_{m}-\alpha_0) \ket{\phi_0}_s
\end{split}
\end{equation}
and the conditional probability of obtaining the outcome $a$ for the system $S$ when measuring the observable $A$ at a certain time $\alpha_{m}$ is given again by the Born rule (see Appendix D):
\begin{equation}\label{born}
\begin{split}
P(a\: on\: S \: |\: \alpha_m \: on \:C) &=\frac{P(a \: on \: S, \: \alpha_m \: on \: C)}{P( \alpha_m \: on \: C)}  =\\ \\&=  \left|	\bra{a}\hat{U}_{s} (\alpha_{m}-\alpha_0) \ket{\phi_0}\right|^2  .
\end{split}
\end{equation}
The Eq. (\ref{born}) shows that the conditioned state of $S$ to a certain clock value $\alpha_{m}$ has no contributions from different times $\alpha_{m'} \ne \alpha_{m}$, and so interference phenomena are not present even if the time states are not orthogonal.

\subsection{Continuous Flow of Time}
\label{continuo}
So far we have considered a discrete flow of time. A continuous can be obtained in the limit $s \to \infty$ \cite{pegg}. 
We define
\begin{equation}
\ket{\tilde{\alpha}} = \sum_{i=0}^{p} e^{- i E_i \alpha}\ket{E_i}
\end{equation}
where again $p+1$ is the number of energy eigenstates and $\alpha$ can now take any real value from $\alpha_0$ to $\alpha_0 + T$. In this framework the resolution of the identity (\ref{pomidentity}) becomes
\begin{equation}\label{newresolution}
\mathbb{1}_c = \frac{1}{T} \int_{\alpha_0}^{\alpha_0+T} d \alpha \ket{\tilde{\alpha}} \bra{\tilde{\alpha}}  .
\end{equation}
The global state is
\begin{equation}\label{applicazioneidentitacontinuo}
\begin{split}
\ket{\Psi} & =  \frac{1}{T} \int_{\alpha_0}^{\alpha_0 + T} d \alpha   \ket{\tilde{\alpha}} \braket{\tilde{\alpha}|\Psi} =\\ \\&= \frac{1}{T} \int_{\alpha_0}^{\alpha_0 + T} d \alpha   \ket{\tilde{\alpha}}_c \otimes \sum_{k=0}^{d_s-1} c_k e^{-i E_k \alpha} \ket{E_k}_s =\\ \\&= \frac{1}{T} \int_{\alpha_0}^{\alpha_0 + T} d \alpha   \ket{\tilde{\alpha}}_c \otimes \ket{\phi(\alpha)}_s 
\end{split}
\end{equation}
and, since $\ket{\phi (\alpha)}_s \propto \braket{\tilde{\alpha}|\Psi}$, we 
derive the Schrödinger equation for the state $\ket{\phi(\alpha)}_s$ 
\begin{equation}
\begin{split}
i \frac{\partial}{\partial \alpha} \ket{\phi(\alpha) }_s &= i \frac{\partial}{\partial \alpha}\braket{\tilde{\alpha}|\Psi}=\\ \\&= i \frac{\partial}{\partial \alpha} \sum_{k=0}^{d_c-1}\bra{E_k}e^{iE_k \alpha}\ket{\Psi} =\\ \\&= - \sum_{k=0}^{d_c-1}\bra{E_k}E_ke^{i E_k \alpha}\ket{\Psi}=\\ \\&=-\bra{\tilde{\alpha}}\hat{H}_c\ket{\Psi}= \hat{H}_s \ket{\phi(\alpha)}_s .
\end{split}
\end{equation}


To conclude this Section we briefly discuss the case of a clock Hamiltonian with a discrete spectrum and arbitrary (not rational) 
energy level ratios. Also in this scenario 
a Schr\"odinger-like evolution is recovered for the state of the system $S$ with respect to the clock. A caveat is that, in this case, the resolution of the identity (\ref{pomidentity}) is no longer exact and time states $\ket{\alpha_{m}}_c$ do not provide an overcomplete basis in $C$. Nevertheless, since any real number can be approximated with arbitrary precision by a ratio between two rational numbers, the residual terms in the resolution of the identity and consequent small corrections can be arbitrarily reduced.


\subsection{Non-Observable Universe as Clock and the Arrow of Time}
\label{discussion}
Is there an arrow of time in the PaW formalism?
The answer in clearly negative. However, it is possible to introduce an {\it emergent} arrow of time. 
Following \cite{vedral} one can consider for simplicity that the system $S$ consists only of two subsystems, \lq\lq the observer\rq\rq (${\Sigma}_1$) and \lq\lq the observed\rq\rq ($\Sigma_2$) which are initially in a product state. The arrow of time (with respect to clock time) can be provided by
the increase in entanglement between the two subsystems within $S$,
as the observer learns more and more about the observed.
We can ask now where to find a good clock for the Universe \cite{arrowoftime}. 
We have considered as a \lq\lq good\rq\rq clock a device defined in a Hilbert space larger than the Hilbert space of the
system $S$, that is $d_c \gg d_s$. Indeed, if $d_c \le d_s$, it would not be possible to connect
every energy state of the system $S$ to an energy state of $C$ satisfying the constraint (\ref{scm}), and some states of $S$ would be excluded from the dynamics. Therefore, the clock introduced here has essentially two properties: 
it has to be larger than the system $S$ and it has to interact only weakly with $S$ or, in the ideal case, it should not interact at all. 
Does such a clock exist? 
A possible choice is to consider the non-observable Universe (namely, the Universe laying outside the light cone centred in the Earth) as a
clock for the observable Universe. Indeed, in this case, the clock and the observable Universe $S$ are not interacting but can still be fully entangled, with the 
Hilbert space of the clock that can be quite larger that the Hilbert space of $S$. With this choice (which of course is just one among several equally speculative choices) the two requirements for a good clock are satisfied. 
We can then consider that during the evolution with respect to such a clock, inside the observable Universe (that is inside the interacting subsystems ${\Sigma}_1$ and ${\Sigma}_2$ of $S$) there is an increasing entanglement generated by $\hat{H}_s$ and, therefore, an increasing relative entropy 
and the emergence of a thermodynamic arrow of time.

\section{Conclusions}
\label{Conclusions}

In this work we have elaborated on the Page and Wootters theory. PaW provides a consistent picture of quantum time as an emerging property of entanglement among subsystems of the Universe. We have considered a protocol introduced by Pegg for the construction of a \lq\lq Age\rq\rq operator complement of a bounded Hamiltonian having a spectrum of equally spaced energy levels. By incorporating Pegg's formalism in the PaW theory we have shown that Age can be interpreted as an Hermitian time operator providing the dynamical evolution of the system.  

In addition we have shown that it is possible to extend this framework to any Hamiltonian with a discrete spectrum having rational ratios of the energy levels. We have demonstrated that even if in this case the time states are not fully distinguishable, the system $S$ still evolves with respect to the clock time according to the Schr\"odinger equation. 
Finally, we have considered the continuous limit of the flow of time.

We can read the PaW approach as a general ``internalization protocol" where, beside time, it is possible to internalise spatial reference frames where space is \lq\lq what is shown on a meter\rq\rq. In the current formalism of quantum mechanics there is an evident asymmetry in how space and time are treated: the spatial degrees of freedom are typically a quantum property of the system under investigation while time on the contrary appears as a classical parameter external to the theory \cite{afundamental}. Perhaps a PaW formulation of spacetime may represent a first step towards removing this asymmetry. This might help to develop relativistic generalizations of the PaW formalism, see also \cite{timedilation,scalarparticles,dirac,ultimamail}.


%

\textit{Acknowledgements} : We acknowledge funding from the project EMPIR-USOQS, EMPIR projects are co-funded by the European Unions Horizon2020 research and innovation programme and the EMPIR Participating States. We also acknowledge financial support from the H2020 QuantERA ERA-NET Cofund in Quantum Technologies projects QCLOCKS.


\bibliographystyle{plain}


\onecolumn\newpage
\appendix

\section{Summary of PaW Theory}
\label{appendicea}
We give here a brief summary of PaW theory, following \cite{vedral}. Page and Wootters consider 
the whole Universe as being in a stationary state with zero eigenvalue (consistently with the the Wheeler-DeWitt equation), that is 

\begin{equation}\label{wdwappendix}
\hat{H}\ket{\Psi} = 0 
\end{equation}
where $\hat{H}$ and $\ket{\Psi}$ are the Hamiltonian and the state of the Universe, respectively. 

They divide the Universe into two non-interacting subsystems, the clock $C$ and the rest of the Universe $S$, and thus the total Hamiltonian can be written as
\begin{equation}\label{h+h1appendix}
\hat{H}=\hat{H}_c\otimes\mathbb{1}_s + \mathbb{1}_c\otimes\hat{H}_s
\end{equation}  
where $\hat{H}_c$ and  $\hat{H}_s$ are the Hamiltonians acting on $C$ and $S$ respectively, and $\mathbb{1}_c$, $\mathbb{1}_s$ are unit operators. 
The condensed history of the system $S$ is written through the entangled global stationary state $\ket{\Psi} \in \mathcal{H} = \mathcal{H}_c \otimes \mathcal{H}_s$ (which satisfies the constraint (\ref{wdwappendix})) as follows:
\begin{equation}\label{statoespansoappendix}
\ket{\Psi} = \sum_{t} c_t \ket{t}_c \otimes \ket{\phi_t}_s 
\end{equation}
where the states $\left\{ \ket{t}_c \right\}$ are eigenstates of the operator choosen to be the clock observable. 

The relative state (in Everett sense \cite{everett}) of the subsystem $S$ with respect to the clock $C$ can be defined

\begin{equation}\label{defstatos}
\rho^{(s)}_t = \frac{Tr_c\left[P^{(c)}_t \rho \right]}{Tr\left[P^{(c)}_t \rho \right]} = \ket{\phi_t}\bra{\phi_t}
\end{equation}
where $\rho = \ket{\Psi}\bra{\Psi}$ and $P^{(c)}_t$ is the projector on a certain time state in the clock subspace. Note that equation (\ref{defstatos}) is the Everett \textit{relative state} definition of the subsystem $S$ with respect to the clock system $C$. As pointed out in \cite{vedral}, this kind of projection has nothing to do with a measurement. Rather, $\ket{\phi_t}_s$ is a state of $S$ conditioned to the clock $C$ being in the state $\ket{t}_c$. 

From equations (\ref{wdwappendix}), (\ref{h+h1appendix}) and (\ref{defstatos}), it is then possible to derive the Schrödinger equation for the relative state of the subsystem $S$ with respect to the clock $C$:

\begin{equation}
\frac{\partial \rho^{(s)}_t}{\partial t} =i \left[\rho^{(s)}_t ,\hat{H}_s \right] .
\end{equation}
Since the subsystem $S$ experiences a Schrödinger-like evolution with respect to the clock $C$, the parameter $t$ can be interpreted as time and the evolution of $S$ has been recovered within a globally stationary Universe. 

The last point concerns conditional probabilities. In the PaW framework the probability to obtain the outcome $a$ when measuring the observable $\hat{A}$ 
on the subspace $S$ \lq\lq at a certain time\rq\rq $\tilde{t}$ can be written as: 
\begin{equation}
\begin{split}
P(a \: on \: S \: | \: \tilde{t} \: on \: C) = \frac{P(a \: on \: S , \: \tilde{t} \: on \: C)}{P( \tilde{t} \: on \: C)}
\end{split}
\end{equation}
that is the conditional probability of obtaining $a$ on $S$ given that the clock $C$ shows $\tilde{t}$. That's why the PaW mechanism is sometimes called \lq\lq conditional probability interpretation of time\rq\rq. 

The PaW approach to time has not been without criticism. For instance Kuchar \cite{kuchar} questioned the possibility of constructing a two-time propagator and Albrecht and Iglesias \cite{iglesias} stressed how the possibility for different choices of the clock inexorably leads to an ambiguity in the dynamics of the rest of the Universe. These objections were addressed by Giovannetti, Lloyd and Maccone \cite{lloydmaccone} (see also \cite{esp2,dolby}) and Marletto and Vedral \cite{vedral}, respectively.

\section{Proof of Equation (\ref{pomidentity})}
\label{appendiceb}
We prove that $\sum_{m=0}^{s}\ket{\alpha_{m}}\bra{\alpha_{m}}$ is equal to the identity:

\begin{equation}\label{dada}
\begin{split}
\sum_{m=0}^{s}\ket{\alpha_{m}}\bra{\alpha_{m}} &= \frac{1}{p+1} \sum_{m=0}^{s}\sum_{i}\sum_{k}e^{-i\alpha_{m}E_{i}}e^{i\alpha_{m}E_k}\ket{E_{i}}\bra{E_k}=\\ \\&= \frac{1}{p+1} \left[   \sum_{m=0}^{s}\sum_{k}\ket{E_{k}}\bra{E_k} +   \sum_{k \ne i}\sum_{m=0}^{s}  e^{i \alpha_{m}(r_k - r_{i})2\pi/T}\ket{E_i} \bra{E_k}  \right] .
\end{split}
\end{equation}
For $(E_k-E_0)/(E_1-E_0)$ rational, and thus $r_k - r_{i}$ integer, the second term of the right side of equation (\ref{dada}) will be zero and then we have

\begin{equation}
\frac{p+1}{s+1} \sum_{m=0}^{s} \ket{\alpha_{m}}\bra{\alpha_{m}} = \mathbb{1}_c .
\end{equation}

\section{Relative State Definition for $S$ in case of non-orthogonal Time States}
\label{appendicec}
We start considering the global state $\ket{\Psi}$ written as

\begin{equation}
\ket{\Psi} = \sum_{k=0}^{d_s-1} \tilde{c}_k \ket{E=- E_k}_c \otimes \ket{E_k}_s
\end{equation}
and we apply in sequence the resolutions of the identity on the clock subspace 

\begin{equation}
\frac{d_c}{D} \sum_{m=0}^{D-1} \ket{\alpha_{m}}\bra{\alpha_{m}} = \mathbb{1}_c
\end{equation}
and 

\begin{equation}
\sum_{n=0}^{d_c-1}\ket{E_n}\bra{E_n}=\mathbb{1}_c.
\end{equation}
We obtain

\begin{equation}
\begin{split}
\ket{\Psi} &= \frac{d_c}{D}\sum_{m=0}^{D-1}\ket{\alpha_{m}}\braket{\alpha_{m}|\Psi}=\\ \\&=\frac{d_c}{D}\sum_{m=0}^{D-1}\ket{\alpha_{m}}_c \otimes \sum_{k=0}^{d_s-1}\tilde{c}_k\braket{\alpha_{m}|E=-E_k} \ket{E_k}_s =\\ \\&= \frac{\sqrt{d_c}}{D}\sum_{m=0}^{D-1}\ket{\alpha_{m}}_c \otimes \sum_{k=0}^{d_s-1}\tilde{c}_k e^{-i\alpha_{m}E_k} \ket{E_k}_s=\\ \\&= \sum_{n=0}^{d_c-1} \ket{E_n}\bra{E_n} \frac{\sqrt{d_c}}{D}\sum_{m=0}^{D-1}\ket{\alpha_{m}}_c \otimes \sum_{k=0}^{d_s-1}\tilde{c}_k e^{-i\alpha_{m}E_k} \ket{E_k}_s=\\ \\&= \frac{\sqrt{d_c}}{D} \sum_{n=0}^{d_c-1} \ket{E_n}_c \otimes \sum_{m=0}^{D-1} \braket{E_n|\alpha_{m}} \sum_{k=0}^{d_s-1} \tilde{c}_k e^{-i\alpha_{m}E_k} \ket{E_k}_s =\\ \\&= \frac{1}{D} \sum_{n=0}^{d_c-1} \ket{E_n}_c \otimes \sum_{m=0}^{D-1} e^{-i\alpha_{m}E_n} \sum_{k=0}^{d_s-1} \tilde{c}_k e^{-i\alpha_{m}E_k} \ket{E_k}_s =\\ \\&= \sum_{n=0}^{d_c-1} \ket{E_n}_c \otimes \sum_{k=0}^{d_s-1}\tilde{c}_k \frac{1}{D} \sum_{m=0}^{D-1} e^{-i \alpha_{m} \left(E_n + E_k\right) } \ket{E_k}_s
\end{split}
\end{equation}
from which we have

\begin{equation}\label{delta}
\sum_{m'=0}^{D-1} e^{-i \alpha_{m'} \left(E_n + E_k\right) } = D\delta_{E_n,-E_k} .
\end{equation}

Considering now the definition of the $\ket{\phi_m}_s$ state as 

\begin{equation}\label{hfhfhfhfhfhfh1}
\ket{\phi_m}_s = \sqrt{d_c} \braket{\alpha_m|\Psi} ,
\end{equation}
we have:

\begin{equation}
\begin{split}
\braket{\alpha_{m}|\Psi} & = \bra{\alpha_{m}} \frac{\sqrt{d_c}}{D} \sum_{m'=0}^{D-1} \ket{\alpha_{m'}} \otimes \sum_{k=0}^{d_s-1} \tilde{c}_k e^{-i\alpha_{m'} E_k} \ket{E_k}_s =\\ \\&= \frac{\sqrt{d_c}}{D} \sum_{m'=0}^{D-1} \sum_{n,n'=0}^{d_c-1} \sum_{k=0}^{d_s-1} \frac{1}{d_c} e^{i E_n \alpha_{m}} e^{-i E_{n'} \alpha_{m'}} \braket{E_n|E_{n'}} \tilde{c}_k e^{-i E_k \alpha_{m'}} \ket{E_k}_s =\\ \\&= \frac{\sqrt{d_c}}{D} \sum_{m'=0}^{D-1} \sum_{n,n'=0}^{d_c-1} \sum_{k=0}^{d_s-1} \frac{1}{d_c} e^{i E_n \alpha_{m}} e^{-i E_{n'} \alpha_{m'}} \delta_{E_n,E_{n'}} \tilde{c}_k e^{-i E_k \alpha_{m'}} \ket{E_k}_s =\\ \\&= \frac{1}{D\sqrt{d_c}} \sum_{n=0}^{d_c-1} e^{iE_n \alpha_{m}} \sum_{k=0}^{d_s-1} \tilde{c}_k \sum_{m'=0}^{D-1} e^{-i \alpha_{m'} \left(E_n + E_k\right)} \ket{E_k}_s ,
\end{split}
\end{equation}
and considering (\ref{delta}) we obtain

\begin{equation}\label{lolol}
\braket{\alpha_{m}|\Psi} = \frac{1}{\sqrt{d_c}} \sum_{k=0}^{d_s-1} \tilde{c}_k e^{-i\alpha_{m} E_k} \ket{E_k}_s.
\end{equation}
Then the definition (\ref{hfhfhfhfhfhfh1}) implies 

\begin{equation}
\ket{\phi_m}_s = \sum_{k=0}^{d_s-1} \tilde{c}_k e^{-i \alpha_{m} E_k} \ket{E_k}_s.
\end{equation}

\section{Proof of Equation (\ref{born})}
\label{appendiced}
We start considering the global state written as

\begin{equation}
\ket{\Psi} = \frac{\sqrt{d_c}}{D} \sum_{m=0}^{D-1} \ket{\alpha_{m}}_c \otimes \ket{\phi_{m}}_s 
\end{equation}
where $\ket{\phi_{m}}_s = \sum_{k=0}^{d_s-1} \tilde{c}_k e^{-i E_k \alpha_{m}} \ket{E_k}_s$. We can now calculate the conditional probability as follows

\begin{equation}\label{ultimadim}
\begin{split}
P(a\: on\: S \: | \: \alpha_m \: on \:C) &=\frac{P(a \: on \: S , \: \alpha_m \: on \: C)}{P( \alpha_m \: on \: C)}  =\\ \\&= \frac{\left| (\bra{\alpha_m}\bra{a})\ket{\Psi}\right|^2}{\sum_{a}\left| (\bra{\alpha_m}\bra{a})\ket{\Psi}\right|^2} =\\ \\&= \frac{\left| (\bra{\alpha_m}\bra{a})\frac{\sqrt{d_c}}{D} \sum_{m'=0}^{D-1} \ket{\alpha_{m'}}_c\ket{\phi_{m'}}_s\right|^2}{\sum_{a}\left| (\bra{\alpha_m}\bra{a})\frac{\sqrt{d_c}}{D} \sum_{m'=0}^{D-1} \ket{\alpha_{m'}}_c \ket{\phi_{m'}}_s\right|^2} =\\ \\&= \frac{\left| (\bra{\alpha_m}\bra{a}) \sum_{m'=0}^{D-1} \ket{\alpha_{m'}}_c\sum_{k=0}^{d_s-1} \tilde{c}_k e^{-i E_k \alpha_{m'}} \ket{E_k}_s\right|^2}{\sum_{a}\left| (\bra{\alpha_m}\bra{a}) \sum_{m'=0}^{D-1} \ket{\alpha_{m'}}_c \sum_{k=0}^{d_s-1} \tilde{c}_k e^{-i E_k \alpha_{m'}} \ket{E_k}_s\right|^2}  =\\ \\&=  \frac{\left| \sum_{m'=0}^{D-1} \sum_{n=0}^{d_c-1} e^{iE_n(\alpha_{m}-\alpha_{m'})}\bra{a}   \sum_{k=0}^{d_s-1} \tilde{c}_k e^{-i E_k \alpha_{m'}} \ket{E_k}_s   \right|^2}{\sum_{a}\left| \sum_{m'=0}^{D-1} \sum_{n=0}^{d_c-1} e^{iE_n(\alpha_{m}-\alpha_{m'})} \bra{a}    \sum_{k=0}^{d_s-1} \tilde{c}_k e^{-i E_k \alpha_{m'}} \ket{E_k}_s   \right|^2} =\\ \\&=  \frac{\left|  \sum_{n=0}^{d_c-1} e^{iE_n\alpha_{m}} \bra{a} \sum_{k=0}^{d_s-1} \tilde{c}_k \sum_{m'=0}^{D-1} e^{-i (E_k+E_n) \alpha_{m'}} \ket{E_k}_s   \right|^2}{\sum_{a}\left| \sum_{n=0}^{d_c-1} e^{iE_n\alpha_{m}} \bra{a} \sum_{k=0}^{d_s-1} \tilde{c}_k \sum_{m'=0}^{D-1} e^{-i (E_k+E_n) \alpha_{m'}} \ket{E_k}_s   \right|^2} .
\end{split}
\end{equation}
Thanks to equation (\ref{delta}), that is $\sum_{m'=0}^{D-1} e^{-i \alpha_{m'} \left(E_n + E_k\right) } = D\delta_{E_n,-E_k}$, we have

\begin{equation}
\begin{split}
P(a\: on\: S \: | \: \alpha_m \: on \:C) &=  \frac{\left| \bra{a}\sum_{k=0}^{d_s-1} \tilde{c}_k e^{-iE_k \alpha_{m}} \ket{E_k}_s   \right|^2}{\sum_{a}\left|  \bra{a}\sum_{k=0}^{d_s-1} \tilde{c}_k e^{-iE_k \alpha_{m}} \ket{E_k}_s  \right|^2} =\\ \\&= \frac{\left| \braket{a|\phi_m}   \right|^2}{\sum_{a}\left|  \braket{a|\phi_m}   \right|^2} = \left| \braket{a|\phi_m}   \right|^2 
\end{split}
\end{equation}
and considering that $\ket{\phi_{m}}_s = e^{-i \hat{H}_s (\alpha_{m}-\alpha_{0})} \ket{\phi_0}_s = \hat{U}_{s} (\alpha_{m}-\alpha_0) \ket{\phi_0}$ we obtain

\begin{equation}
P(a\: on\: S \: |\: \alpha_m \: on \:C) = \left|	\bra{a}\hat{U}_{s} (\alpha_{m}-\alpha_0) \ket{\phi_0}\right|^2 .
\end{equation}

\end{document}